\documentclass[%
reprint,
showkeys,
showpacs,preprintnumbers,
amsmath,amssymb,
aps,
pra,
floatfix,
]{revtex4-1}
\usepackage{graphics}
\usepackage{epsfig}
\usepackage{epstopdf}
\usepackage{amsmath,amssymb,stackengine}
\usepackage{newtxmath,newtxtext}
\usepackage{array}
\usepackage{subfigure}
\usepackage[colorlinks=true,linkcolor=blue,citecolor=blue,      urlcolor=blue]{hyperref}
\usepackage{color,soul}

\begin{document}
\title{Unified formalism for the law of emergence from the first law of thermodynamics}
	\author{Hassan Basari V. T.}
	\email{basari@cusat.ac.in}
	\author{P. B. Krishna}
	\email{krishnapb@cusat.ac.in}
	\author{Titus K. Mathew.}
	\email{titus@cusat.ac.in}
	\affiliation{%
		Department of Physics, Cochin University of Science and Technology, Kochi, Kerala 682022, India
		}%
\begin{center}
\begin{abstract}
	We derive a unified expansion law for our universe from the first law of thermodynamics on the apparent horizon, where entropic evolution depicts the emergence of cosmic space. The derivation advances a general form for degrees of freedom on the surface and bulk, which provides a natural generalization for the expansion law proposed by Padmanabhan. The general expression for the surface degrees of freedom differs from the natural expectation, $ N_{sur }= 4S $ in the emergent gravity paradigm for general theories of gravity. The derivation also provides justification for the selection of Gibbons-Hawking temperature in the original expansion law and for the use of areal volume in the non-flat FRW universe. Since the unified expansion law exclusively depends on the form of entropy, the method is applicable to obtain the expansion law in 
	 any gravity theory without any additional ad hoc assumptions. From the general expansion law, we have obtained the expansion law corresponding to different theories of gravity like (n+1) Einstein, Gauss-Bonnet, Lovelock, and Horava-Lifshitz. We also obtained the expansion law for non-extensive entropy like Tsallis entropy from the unified expansion law.

\end{abstract}  
\maketitle
\end{center}
\section{Introduction}
The connection between gravity and thermodynamics
 first emerged from the studies on black hole physics
\cite{bekenstein1,bekenstein2,bardeen,Hawking1}. In 1973, Bekenstein proposed that black holes possess entropy proportional to their horizon area \cite{bekenstein1}. Meantime, Bardeen et al. formulated the laws
of black hole dynamics and have shown that they are analogous to the laws of thermodynamics of an ordinary macroscopic system \cite{bardeen}. Following these, Hawking has revealed that the black
hole possesses temperature proportional to the surface gravity of its horizon; hence, it can radiate like any thermal object. Gibbons and Hawking \cite{PhysRevD.15.2738} extended this idea to cosmology
and showed that the horizon of a de Sitter universe, in-
deed possesses a temperature and entropy proportional to
its surface gravity and area, respectively. Later on, it was noticed that the thermodynamic nature is not a peculiar
feature of the black holes or cosmological horizons but a general feature of
all spacetime horizons \cite{PhysRevD.7.2850, Davies_1975,PhysRevD.15.2738}. For instance, an accelerating observer in flat spacetime can experience a horizon and attribute a
temperature (called Unruh temperature) to it,
$T = a/2\pi,$ where $ a $ is his acceleration \cite{PhysRevD.14.870}. 

In 1995, in an important step,  Jacobson\cite{1995} showed that Einstein's field
equations can be derived by projecting the Clausius relation, $\delta Q = T dS$ to a local Rindler horizon, where $\delta Q$
is the energy flux through the horizon, $T$ is the Unruh
temperature \cite{PhysRevD.14.870} seen by the accelerated observer near
the horizon and $ S $ is the Bekenstein entropy \cite{bekenstein1}. Later,
Padmanabhan had shown that Einstein's field equation
can be expressed in the form $dE = \mathcal{T} dS - P dV,$ near a spherically
symmetric horizon, where $E$ is the energy associated with
the horizon, $P$ is the pressure term due to the gravitational
source (which is different from conventional pressure),
and $\mathcal{T}$ is horizon temperature proportional to the surface gravity \cite{Padmanabhan:2002sha, PhysRevD.74.104015}. The differences between these two approaches are discussed in reference \cite{PhysRevD.83.024026}.

In cosmology, Cai and Kim\cite{Cai_Kim} derived the Friedmann equations from the first law of thermodynamics of the form  $-dE_{_{flux}} = T dS$ by applying it to the horizon of an (n+1)-dimensional FRW universe in Einstein's
gravity. Here $S$ is the Bekenstein entropy of the horizon, $T = 1/2\pi r_A$ is the static temperature of the horizon \cite{Cai:TemperaturApparentHorizon2009}
and $dE_{_{flux}},$ is the energy flux through
the apparent horizon of radius $r_A$. This result has 
been extended to Gauss-Bonnet gravity and the more
general Lovelock gravity theory by adopting the corresponding entropy relations. Meanwhile, Akbar and Cai
arrived at the Friedmann equations from the thermodynamic identity, $dE = \mathcal{T} dS + W dV$ \cite{PhysRevD.75.084003}. Here dE is
the distortion in the Misner-Sharp energy of the matter
inside the apparent horizon, and temperature is assumed as
$\mathcal{T} = \kappa/2\pi,$ where $\kappa$ is the surface gravity of the horizon,
and $W =(1/2)(\rho-p),$ the work density.

All the results discussed above ratifies the intriguing connection between gravity and thermodynamics. This implies that, like thermodynamics, gravity could also be an emergent phenomenon. In thermodynamics, properties like temperature, pressure, etc., emerged due to the assembly of microscopic structures like atoms or molecules. Like this, the macroscopic geometric
properties of spacetime like metric, curvature, etc.  are
also turn out to be emerging entities. Following this concept,
Padmanabhan \cite{paddy2010dec} derived Newton's law of gravity by
combining the equipartition law of energy, related to the degrees
of freedom at the horizon, and the thermodynamic relation, $S = E/2\mathcal{T},$ where $E$ is the effective gravitational
mass and $\mathcal{T}$ is the horizon temperature. Meantime, following string theory considerations, Verlinde reformulated gravity as an entropic force arising from 
the natural tendency of the material distribution to maximize
the entropy \cite{Verlinde}.
 
In the emergent gravity approach, spacetime is considered to
be pre-existing. In extending this approach to cosmology, Padmanabhan further stated that the cosmological space itself could be emergent. He postulated that the expansion of the universe (expansion of the Hubble volume) could be explained as the emergence
of cosmic space with the progress of cosmic time\cite{paddy2012jun}.
It is difficult to assume time as
being emerged from some pregeometric variables. However, the existence of proper time will help to eliminate this difficulty in cosmology.    In cosmology, due to co-moving coordinates, all inertial observers measure the same time, the proper time. Moreover, for such inertial observers, the cosmic background radiation appears homogeneous and isotropic \cite{paddy2012jun}. Padmanabhan then proposed, in the context of Einstein's gravity,  that
the time evolution of the universe can be described using
the equation,
$\dfrac{dV}{dt}= \ell_{p}^{2} \left( N_{sur} -\epsilon N_{bulk} \right),$ 
where $V$ is the Hubble volume, $\ell_p$ is the Planck length and $N_{sur} \& N_{bulk}$
are the degrees of freedom on the horizon and that of
the bulk, residing within the horizon respectively. The
above relation is known as the holographic equipartition
principle and is also dubbed as the expansion law. According to this law, the emergence of space happens to
equalize the degrees of freedom (DoF ) on the horizon
with that in bulk matter enclosed by the horizon.
To establish this concept, Padmanabhan derived the Friedmann equation for a flat
(3+1) FRW universe from this law of expansion \cite{paddy2012jun}.
The expansion law was then extended to higher-dimensional
gravity theories like (n+1) dimensional Einstein gravity,
Gauss-Bonnet gravity and more general Lovelock gravity
by appropriately modifying the surface degrees of freedom \cite{cai}. An extension of this procedure to non-flat
FRW universe was done by Sheykhi \cite{Sheykhi2013}. The expansion law had also been extended to different gravity theories in different ways \cite{cai,Sheykhi2013,yang2012,Yuan:2016pkz,FARAGALI,FLDezaki}. More investigations on Padmanabhan's idea of emergence of space can
be found in references \cite{Padmanabhan2019review,paddycosmologicalconstant, PhysRevD.90.124017,SHEYKHI2018118,TU2018411,PhysRevD.99.043523,PhysRevD.88.084019,Tu_2013,FARAGALI, Yuan:2016pkz,basari1,Mahith2018, Krishna2022}. In recent
studies, we have shown that the expansion law effectively
implies the entropy maximization in Einstein's gravity \cite{krishna1} and more general forms of gravity like Gauss-Bonnet
and Lovelock gravities \cite{krishna2}. Consequently, one can interpret the emergence of space as a tendency to maximize
the horizon entropy.
 
There are different generalizations of Padmanabhan's original proposal of the law of emergence in the context of different gravity theories. In these various generalizations, the authors have to define an effective area for the apparent horizon and, thus, a form for the effective degrees of freedom on the horizon, in conformity with the expression for the entropy of the horizon. Owing to the different forms of the Horizon entropy in different gravity theories, the effective surface (or volume) and the surface degrees of freedom defined by different authors vary from one gravity theory to another. This ultimately led to the proposal of different forms of the law of emergence in different gravity theories. The authors only made an effort to guarantee the emergence of the correct Friedmann equations, whatever way the form of the law of emergence varies for different gravity theories. Hence the situation cried for having a unified format for the law of emergence, irrespective of the gravity theory. It is to be noted that the different forms of the expansion law can be derived from the first law of thermodynamics, which has the same form in all gravity theories \cite{Mahith2018,FLDezaki}. So we think it is possible to find a unified general form for the law of emergence that can be invariantly applicable in all gravity theories, from the principles of thermodynamics. Motivated by this, we formulate a more general and unified form for the law of emergence in terms of entropy, which can be applied to any gravity theory with its unique entropy relation. 

The paper is organized as follows. In section \ref{sec.2},  
we derive the general expansion law from the unified first law of thermodynamics. Following this, we obtain expansion law in different theories of gravity from the general expansion law in section \ref{sec.3}. The significance of the unified expansion law is discussed in Sec. \ref{sec.significance}. Finally, we present our conclusions in section \ref{sec.4}.

\section{General expansion law from unified First law of thermodynamics} \label{sec.2}
According to Padmanabhan, the dynamics of the universe can be explained as the emergence of space by the expansion law \cite{paddy2012jun}. Similarly, the unified first law relates the variation in Misner-Sharp energy with the heat-supply term $\mathcal{T}dS$ and the work term $WdV$ due to the variation in the horizon radius. Both relations explain the same reality of the dynamics of the universe. Hence there should exist deeper connections between them. In this section, we derive expansion law from the unified first law of thermodynamics. The derivation shows that the entropy determines the nature of gravity in the resulting expansion law as in the unified first law. Hence the derivation gives a general form of the expansion law that allows us to transcend the idea of emergence to general gravity theories like Gauss-Bonnet and Lovelock gravity theories by choosing respective entropies. The derivation also gives insights for defining the surface degrees of freedom in different gravity theories.  

Let us consider an (n+1) dimensional FRW universe with metric 
\begin{equation}\label{eq:metric}
ds^{2} = h_{ab}dx^{a}dx^{b} + a(t)^{2}r^{2}d\Omega_{n-1}^{2},
\end{equation}
where $h_{ab}=\text{ diag}\left[-1, a(t)^2/1-kr^2\right]$ is the two dimensional metric 
of the  $ t-r $ surface, $a(t)$ is the scale factor of  expansion, $r$ is the co-moving radial distance, and $d\Omega_{n-1}$ is the metric of (n-1)-dimensional sphere with unit radius.  The spatial curvature constant have values 
$k = 1, 0 \, \textrm{and} \, -1,$ corresponding to a closed, flat and open universe and, $a(t)$ is the scale of expansion. The apparent horizon of the universe satisfies the condition,
$h_{ab} \partial_a\tilde{r} \partial_b\tilde{r}=0,$ (where $\tilde{r}(t)= a(t)r$), which gives  the apparent horizon radius as, $  \tilde{r}_{_A} = 1/\sqrt{H^2 + (k/a^2)} $,
where $H$ is the Hubble parameter. From the standard relation for the surface gravity $\kappa$, the horizon temperature have the form\cite{Hayward_1998,Hayward:1998ee},
\begin{equation}\label{eq:temperature}
 \mathcal{T} = -\kappa /2\pi = \frac{1}{2\pi \tilde{r}_{_A}}  \left(1-\epsilon _{_T}\right)=T \left(1-\epsilon _{_T}\right)
\end{equation}  
where $\epsilon_{_T} \equiv \dot{\tilde{r}}_{_A}/2H\tilde{r}_{_A} $, with the over-dot represents a derivative with respect to cosmic time 
and $T=1/2\pi  \tilde{r}_{_A}.$ In the limit $\dot{\tilde{r}}_{_A}/2H\tilde{r}_{_A} << 1$ the temperature satisfies the relation $\mathcal{T} \sim T$\cite{Cai:TemperaturApparentHorizon2009}. Here the negative sign before $\kappa$ is introduced to ensure the non-negativity of the temperature. In the cosmological context, unlike in the blackhole case, the surface gravity of the horizon, $\kappa \le 0$ for the standard cosmic components $\omega \le 1/3$ \cite{DavidIvanBooth}. 

Now, the unified first law  can be expressed as
\begin{equation}\label{eq:unifiedFirstLaw}
dE=-\mathcal{T} dS+WdV   ,
\end{equation}
where the negative sign in $TdS$ cancels out with the negative sign in the definition of temperature, $\mathcal{T}=-\kappa/2\pi$, 
see \cite{Padmanabhan:2002sha} for more about the sign conventions.  We use the above form of the unified first law in the rest of this manuscript. In cosmology, 
there also have another form of the first law of thermodynamics 
without pressure term, as $-dE_{_{flux}}=T dS$ \cite{Cai_Kim}. The unified first law (\ref{eq:unifiedFirstLaw}) is applied to the entire volume within the horizon, in which 
the energy is the Misner-Sharp energy, $E=\rho V$ contained within the horizon of volume $V$ \cite{PhysRevD.75.084003}. On the other hand, $-dE_{_{flux}}=TdS$ is used at the horizon, and the energy $dE_{flux}$ is referred to as the energy flux crossing the apparent horizon in an infinitesimal interval of time during which the size of the horizon is assumed to be fixed, and hence the temperature of the horizon is assumed to be $T= 1/2\pi r_{_A}$ \cite{Cai_Kim}. 

Let us now formulate the unified first law at the apparent horizon in the FRW universe (\ref{eq:metric}). 
The cosmic component is assumed to be a 
perfect fluid, such that the time and spacial components of the energy-momentum tensor are, $T^{0}_{0}=-\rho;~T^{i}_{i}= p$ with density $\rho$ and pressure $p$ of the cosmic components,
\begin{equation}\label{eq:Tmunu}
T_{\mu \nu}=  \left(\rho + p\right)u_{\mu}u_{\nu} + g_{\mu \nu}p.
\end{equation}

 Thus the energy within the volume $V$ enclosed by the apparent horizon is $E=\rho V.$ Then, the unified first 
law can be expressed as \cite{PhysRevD.75.084003},
\begin{equation}
\mathcal{T}dS = \frac{\left(\rho - p\right)}{2}dV - \left(\rho dV + Vd\rho\right) .
\end{equation}
Using the continuity equation in (n+1) FRW universe, $\dot{\rho} + nH\left(\rho +p \right)$ = 0, the above equation becomes
\begin{equation}
\label{eq:ds/dt} 
T\dfrac{dS}{dt} = nH\Omega_{n}\tilde{r}_{_A}^{n}\left(\rho + p\right).
\end{equation} 
Here the temperature is $T,$ since the term $\left(1-\epsilon _{_T}\right) $ was cancelled, from both sides.
In obtaining the above relation,  
we took $V = \Omega_{n}\tilde{r}_{_A}^{n}$, the volume of (n+1) FRW universe enclosed by the apparent horizon, where $\Omega_{n}$ is the areal volume of an n-dimensional sphere with unit radius. 

It is important to note that the first law of the form, $-dE_{_{flux}} =T dS$ also reduces Eq. (\ref{eq:ds/dt}).
 It should be noted that, unlike in the previous form of the law, the energy $dE_{_{flux}}$ here is 
the energy flux across the apparent horizon. The observer measuring this flux is located on the apparent horizon,  
for whom 
the apparent horizon is virtually stationary.  
Relative to this local observer, the temperature of the apparent horizon becomes $T=1/2\pi \tilde{r}_{_A} $.  
Energy flux through the apparent horizon 
during a small interval of time, $dt$ is given by \cite{AKBAR20067} 
\begin{equation}
-dE_{flux} =  
A\left(\rho+p\right)H\tilde{r}_{_A} dt,
\end{equation}
where $A= n\Omega_{n}\tilde{r}_{_A}^{n-1}$ is the area of the horizon of an (n+1) dimensional FRW universe.
Then the first law 
at the apparent horizon of the FRW universe will take the form,
\begin{equation}
T dS = A\left(\rho+p\right)H\tilde{r}_{_A} dt.
\end{equation}
On substituting the area of the apparent horizon with some suitable rearrangements, the above equation becomes exactly similar to 
Eq. (\ref{eq:ds/dt}), obtained for the unified form of the law of thermodynamics. So that Eq. (\ref{eq:ds/dt}) represents both forms of the thermodynamic laws at the horizon. 

Let us now consider Eq. (\ref{eq:ds/dt}) and split the term $ n(\rho +p) $ on the R. H. S. of it  into $ (n-2)\rho +np +2\rho.$ Using this, Eq. (\ref{eq:ds/dt}) can be rewritten as,
\begin{equation}
\label{eq:ds/dt1}
\frac{dS}{dt} =  \frac{1}{T}H\Omega_{n}\tilde{r}_{_A}^{n} \left[\left(n-2\right)\rho + np+  2 \rho \right].
\end{equation}
Now we take the rate of change of horizon entropy, $dS/dt$, instead of the rate of change of volume enclosed by the apparent horizon, $dV/dt$, in the expansion law. This is useful in generalizing the expansion law to more general theories of gravity having different forms of entropy. Similarly, we use a natural generalization of Komar energy to define the degrees of freedom in bulk for (n+1) FRW universe, $ E_{\!_{Komar}}= 2\left[(n-2)\rho+np \right] V/(n-1). $ It turns out that, this relation for Komar energy can be obtained from its standard relation \cite{PhysRevD.90.124017}, 
\begin{equation} \label{eq:KomarEnergy}
E_{Komar} = 2\int dV \left(T_{\mu \nu}-\frac{1}{(n-1)}\bar{T}g_{\mu\nu}\right)u^{\mu}u^{\nu} ,
\end{equation}
where  $T_{\mu\nu}$ is the energy-momentum tensor, with trace, $\bar{T}$ and $u^{\mu}$ is the four velocity. For (n+1) FRW universe, the energy-momentum tensor given in equation (\ref{eq:Tmunu}) satisfies the relations, $T_{\mu\nu}u^{\mu}u^{\nu}= \rho$ and  $\bar{T} g_{\mu\nu}u^{\mu}u^{\nu}= \rho +np.$ Using these it is easy to show that, equation (\ref{eq:KomarEnergy}) will reduce to the Komar energy that we have used to define bulk degrees of freedom. Taking account of these two assumptions, with suitable rearrangements in Eq. (\ref{eq:ds/dt1}), we get a general expansion law \footnote{A similar relation is expressed in \cite{Krishna2022}. The aim of the current results is different from these works. This is explained in sec. \ref{sec.comparison}}, 
\begin{equation}\label{eq:expansionlaw}
\frac{4}{(n-1)}\dfrac{dS}{dt} = H\left(N_{sur}  -\epsilon N_{bulk}\right),
\end{equation}
where 
\begin{equation}\label{eq:Nbulk}
N_{bulk}= -\epsilon \frac{E_{\!_{Komar}}}{(1/2)T}=-\epsilon \left(\frac{2\left[(n-2)\rho+np \right] V}{(n-1)}\right)\frac{1}{(1/2)T}.
\end{equation}
Here we can identify the $N_{sur}$ as the total surface degrees of freedom for a general (n+1) dimensional FRW universe, and it turns out that  
	\begin{equation}\label{N_sur}
	N_{sur}= \frac{2}{n-1}\frac{2\rho V }{(1/2)T},
	\end{equation}
where $n$ is the spacial dimension of the space. In this process, we haven't used any particular gravity theory; hence the above results are true irrespective of the gravity theory. The above relation can be effectively rewritten as $N_{sur}=E_{eq}/((1/2)T)$, where $E_{eq}=\frac{2}{(n-1)}2\rho V$, which can be identified as equipartition energy of the horizon of an (n+1) dimensional universe.
	
In the context of (3+1) Einstein's gravity, $\rho V $ has the direct thermodynamic interpretation as the total heat content of the horizon surface, which is in the form $TS$ \cite{paddy2010dec}. Then the present form of $E_{eq}$ will reduce to $E_{eq} = 2TS,$ corresponding to which, $N_{sur}=4S =\frac{A}{\ell_{p}^{2}}$ will be followed \cite{Padmanabhan:2003pk, Padmanabhan:2009kr}. Similarly, the first Friedmann equation in (n+1) Einstein's gravity has the form, $ TS=\frac{2}{(n-1)} \rho V $. Hence, our form of the equipartition energy in this case also reduces to $E_{eq} =\frac{2}{(n-1)} 2\rho V= 2TS $, correspondingly, $N_{sur}=4S=\frac{A}{\ell_{p}^{n-1}}  $. 
 
However, in general theories of gravity, especially in Lovelock gravity theory, the entropy is not proportional to the area of the horizon, which no longer follows such a simple relation $\frac{2}{(n-1)}\rho V = TS$ \cite{PhysRevD.75.084003}, and corresponding surface degrees of freedom will not follow the relation $N_{sur} = 4S$. Here, one can note that the form of equipartition energy on the horizon determines the degrees of freedom on the surface. Now we take the general assumption that the equipartition energy of the surface is two times the heat content of the horizon \cite{2014GReGr}. Then, $\frac{2}{n-1} \rho V$ in the above relation should be equivalent to the total heat content of the horizon (which is not equal to $TS$ in general). This can be realized using the Clausius relation $TdS= -dE_{flux}= -Vd\rho$, which means the change in the total heat content on the horizon surface is equal to the total inward matter flux through the horizon for a local observer \cite{Cai_Kim}. Using the integral form of the Clausius relation, we get the equipartition energy of horizon as 
	\begin{equation}
	E_{eq} = \frac{4}{n-1} \rho V = -\frac{4}{n-1} V\int \frac{TdS}{V}.
	\end{equation}
	Hence we identified the equipartition energy in our derivation in terms of thermodynamic variables. So, we get the general expression for the surface degrees of freedom as 
	\begin{equation}\label{eq:Nsur} 
	N_{sur}= \frac{E_{eq}}{(1/2)T} = -\frac{4}{(n-1)}\frac{V}{(1/2)T} \int \frac{TdS}{V}.
	\end{equation} 

 Now, we want to highlight that equation (\ref{eq:expansionlaw}) is the general law of expansion, and equations (\ref{eq:Nsur}) and (\ref{eq:Nbulk}) are the surface and bulk degrees of freedoms, respectively.
 First, we will show that our generalized expansion law in equation (\ref{eq:expansionlaw}) will reduce to the original form of the law of expansion, which was proposed for the (3+1) flat FRW universe in the context of Einstein's gravity. For this, let us take, $n=3$ and entropy as, $S=A/4\ell_p^2,$ with $A=4\pi/H^2$ for a flat universe. Then the, left hand side of equation (\ref{eq:expansionlaw}) will reduce to $(H/\ell_p^2) \frac{dV}{dt},$ where $V=4\pi/3H^3$ , the volume of the apparent horizon. To find the surface degrees of freedom, we have to substitute $A, V$ and $T=H/2\pi,$ the temperature for the horizon, in equation (\ref{eq:Nsur}), which on integration will lead to $N_{sur}=A/\ell_p^{2}.$  Similarly, the bulk degrees of freedom in equation (\ref{eq:Nbulk}), will now reduce to $N_{bulk}=(\rho+3p)V/(1/2)T,$ where $(\rho+3p)$ is the Komar energy density for a flat (3+1) universe. Taking account of these, the expansion law in (\ref{eq:expansionlaw}) will now reduce to the form, 
\begin{equation}
\frac{dV}{dt}=\ell_p^2 \left(N_{sur}-\epsilon N_{bulk} \right),
\end{equation}
which is exactly the original form of the law of expansion of the flat universe. 

The novel fact regarding the general law of expansion is the symmetry in the definitions of  $N_{sur}$ and $N_{bulk}.$  In the original proposal by Padmanabhan, $N_{sur}$ is defined as the number of Planck area, which corresponds to a degree of freedom on the horizon surface, while $N_{bulk}$ is defined by using the equipartition rule, as the ratio of the Komar energy within the horizon to the energy corresponding to one degree of freedom. In the current approach, an alternative definition for the surface degrees of freedom, $N_{sur}$ similar to $N_{bulk}$ can be obtained using the equipartition rule.  
 It turns out that the expression, $-\frac{4V}{(n-1)} \int \frac{TdS}{V}$ in equation (\ref{eq:Nsur}) is equivalent to 
  equipartition energy of the surface, $E_{eq}, $ so that the surface degrees of freedom can consequently be expressed as, $N_{sur}=\frac{E_{eq}}{(1/2)T}.$ This brings symmetry to the definition of the bulk degrees of freedom, which takes the form $N_{bulk}=-\epsilon \frac{E_{Komar}}{(1/2)T}.$ However, it is to be noted that the temperature used to define both degrees of freedoms is the same, $T =1/(2\pi \tilde{r}_{_A})$. This implies that the bulk degrees of freedom that emerged in bulk are in equipartition with the horizon at temperature $T $  as in the original proposal by Padmanabhan.

An important point about the expansion law (\ref{eq:expansionlaw}) is regarding the temperature appearing in this equation. To derive the
expansion law from the unified first law, we have used the temperature, $\mathcal{T}$ as defined in equation (\ref{eq:temperature}). This temperature is a measure of the surface gravity of the horizon and is a product
of two terms, in which the first term is proportional to $1/\tilde{r}_A$,
and the other mainly depends on ${\dot{\tilde{r}}}_A,$ the time derivative of the
horizon radius. It is to be noted that many have used the
horizon temperature with this time derivative of the radius as
the temperature of the dynamic horizon. This motivates the
use of the temperature $\mathcal{T}$ to define the expansion law, especially for obtaining the degrees of freedom. However, in the
original proposal of the law of emergence, by Padmanabhan,
the temperature used is $T$, which is devoid of the time derivative of the horizon radius. The justification given for this is
threefold: (1) the temperature  $T$ is independent of the gravity
theory, and also, it is a direct translation of the expression of
the temperature of a black hole horizon; (2) since $r_A$ is also
varying as the universe expands, it also accounts the dynamical
evolution of the universe; (3) the expansion law obtained with
the temperature $T,$  ($T=H/2\pi$ for the flat universe, used by
Padmanabhan) has a simple form, which will reduce to the
Friedmann equation in the case of a flat universe. The interesting point in our calculation process is that, even though we
have started with the full temperature $\mathcal{T},$ the final expression
for the expansion law contains only $T$, which is arisen so naturally. In a sense, the use of the temperature, $T$ for defining the
expansion law has a natural explanation in our method.

  \section{ Expansion law in different theories of gravity from the general law of expansion}\label{sec.3}
 In this section, we will obtain the law of expansion for different gravity theories from the general law of expansion presented in the last section. We will then compare these with the existing generalizations of the law. 
   
 \subsection{Expansion law in (n+1) Einstein gravity} 
  The entropy of the horizon in (n+1) dimensional Einstein gravity is, $S= A/(4\ell_{p}^{n-1})$ and its time derivative can be expressed as, $ \frac{(n-1)}{4\ell_{p}^{n-1}\tilde{r}_{_A}}\frac{dV}{dt} $. Using this, the general expansion law (\ref{eq:expansionlaw}) will reduce to the form,
  \begin{equation}\label{eq:EinsteinExpansionLaw}
  \frac{dV}{dt}=\ell_p^{n-1} \tilde{r}_A H \left(N_{sur}-\epsilon N_{bulk}\right).
  \end{equation}
  This is the law of expansion in (n+1) dimensional Einstein's gravity. For the flat universe with the apparent horizon radius, $\tilde{r}_{\!\!_A}=1/H$, the expansion law will reduce to
  \begin{equation}\label{eq:EinsteinExpansionLawflat}
  \frac{dV}{dt}=\ell_p^{n-1}\left(N_{sur}-\epsilon N_{bulk}\right).
  \end{equation}
  This equation is of the same form as the original proposal by Padmanabhan.  
  The degrees of freedom appearing in the above equation can be obtained from the general expressions in $ (\ref{eq:Nsur}) $ and $ (\ref{eq:Nbulk}) $, now take the form,  
  \begin{equation}\label{EinsteinDoF}
  N_{sur} = A/\ell_{p}^{n-1} \text{ and } N_{bulk}= -\epsilon \frac{E_{_{Komar}}}{(1/2)T}.
  \end{equation}
  
  We will now show that the expansion law (\ref{eq:EinsteinExpansionLaw}) will lead to the corresponding Friedmann equation. Substitute for $V, A, E_{_{Komar}}$ and $T$ in the expansion law, equation (\ref{eq:EinsteinExpansionLaw}) will be modified as,
  \begin{equation}
  \tilde{r}_{\!\!_A}^{-2} - \dot{\tilde{r}}_{\!\!_A}H^{-1}\tilde{r}_{\!\!_A}^{-3} =\frac{8\pi \ell_{\! \!_p}^{n-1}}{n(n-1)} \left[(n-2)\rho + np\right].
  \end{equation}
  Integrating the above equation using the continuity equation, $\dot{\rho} + nH (\rho + p)=0$, after multiplying both sides by the factor $2\dot{a}a$, we get the first Friedmann equation\cite{Cai_Kim},
  \begin{equation}
  H^{2} +\frac{k}{a^2} = \frac{16\pi\ell_{\!\!_p}^{n-1}}{n(n-1)}\rho.
  \end{equation}
  This shows the consistency of our expansion law in (n+1) Einstein gravity for both flat and non-flat FRW universe.

Now we will compare the above results with the previously obtained extensions of the law of expansion in (n+1) dimension. The original proposal of the law of expansion, for flat (3+1) Einstein's gravity has extended to (n+1) dimensions in flat FRW universe by R. G. Cai \cite{cai}, while, Sheykhi extended the law to the non-flat universe  \cite{Sheykhi2013}. We will briefly describe Cai's extension, then highlight, how our general expansion law is more advantageous. 

In extending the law of emergence to (n+1) Einstein's gravity, Cai proposed a modified form for the law as\cite{cai}, $\alpha \frac{dV}{dt} = \ell_p^{n-1} \left(N_{sur} - N_{bulk}\right),$ where $\alpha=(n-1)/2(n-2).$ In this  case the degrees of freedom has been taken to be, $N_{sur}=\alpha A/\ell_p^{n-1}$ and $N_{bulk}= E_{\!_{Komar}}/[(1/2)(H/2\pi)],$ with Komar energy, assumed to be, $E_{Komar}=\left[((n-2)\rho+np)/(n-2)\right]V,$ where $V$ is the volume of the Hubble horizon in (n+1) dimension. In general, Cai's extension has the following features, contrary to the original proposal:
\begin{enumerate}
	\item The appearance of $\alpha$ on the left hand side of this equation makes it different in form compared to the original proposal.   
	\item There contains an additional coefficient $\alpha,$ contrary to the original proposal, in the definition of the surface degrees of freedom, as $N_{sur}=\alpha A/\ell_p^{n-1}$.  Hence it deviates from the original rule that $N_{sur}=4S.$
	\item Similarly, the Komar energy, $ ((n-2)\rho+np)V/(n-2) $ is not in conformity with the one following from the standard expression as given in equation (\ref{eq:KomarEnergy}), there have an additional coefficient $\alpha$ in the Cai's definition for Komar energy.                 
\end{enumerate}
Consequently, both the surface and bulk degrees of freedom and the L. H. S. of the expansion law in reference \cite{cai,Sheykhi2013} have an additional coefficient $\alpha$ for $n\ge 4$.  This implies that the minimum area in (n+1) dimensional universe is assumed as $\alpha^{-1}\ell_{\!\!_p}^{n-1}$ \cite{Verlinde,Chang-Young:2013gwa}, while it is mere $\ell_{\!\!_p}^{n-1}$ in our definition, which is more simple that agrees with the definition in \cite{PhysRevD.90.124017}. So we can say that the law of expansion for (n+1) Einstein's gravity, which arises from our general law, is exactly a straightforward extension of the original proposal.

\subsection{Expansion law in Lovelock and Gauss-Bonnet theories of gravity}

Now we deduce the expansion law for more general theories of gravity like Gauss-Bonnet and Lovelock gravity from the general expansion law. In 
Lovelock gravity the entropy has the form \cite{CAI2004237,Sheykhi2013},
\begin{equation}\label{eq:entropy_lovlock}
S = \dfrac{A}{4\ell_{p}^{n-1}}\sum_{i=1}^{m}\dfrac{i(n-1)}{(n-2i+1)}\hat{c}_{i}\tilde{r}_{_A}^{2-2i},
\end{equation}
which has an additional correction term in comparison with that of Einstein's gravity. Consequently, the time derivative of above entropy is $\frac{(n-1)}{4\ell_{p}^{n-1}\tilde{r}_{_A}}\frac{dV}{dt} \sum_{i=1}^{m}i\hat{c}_{i}\tilde{r}_{_A}^{2-2i} .$ 
Using this, the general expansion law in Lovelock gravity can be written as
\begin{equation} \label{eq: expansionlaw_lovlock}
\frac{dV}{dt} = \frac{\ell_{p}^{n-1}\tilde{r}_{_A}H}{\sum_{i=1}^{m}i\hat{c}_{i}\tilde{r}_{_A}^{2-2i}}\left[ N_{sur} - \epsilon N_{bulk} \right].
\end{equation} 
Further, the surface degrees of freedom can be obtained from the general expression (\ref{eq:Nsur}) as 
\begin{align}\label{eq:nsur_lovlock}
N_{sur} &= -\frac{2}{(n-1)}4 \tilde{r}_{_A}^{n+1}\int \frac{1}{\tilde{r}_{_A}^{n+1}}dS 
\nonumber\\
&= \frac{A}{\ell_{p}^{n-1}}\sum_{i=1}^{m}\hat{c}_{i}\tilde{r}_{_A}^{2-2i}.
\end{align} 
And the bulk degrees of freedom as in eq. (\ref{eq:Nbulk}).   

Now we will show that the expansion law we obtained in Lovelock gravity will give the corresponding Friedmann equation. Substitute for $V, A, E_{_{Komar}}$ and $T$ in the expansion law, we then get equation (\ref{eq: expansionlaw_lovlock}) as,
\begin{equation}
\sum_{i=1}^{m}\hat{c}_{i}\tilde{r}_{\!\!_A}^{-2i} - \dot{\tilde{r}}_{\!\!_A}H^{-1}\sum_{i=1}^{m}i\hat{c}_{i}\tilde{r}_{\!\!_A}^{-2i-1} =\frac{8\pi \ell_{\! \!_p}^{n-1}}{n(n-1)} \left[(n-2)\rho + np\right].
\end{equation}  
We first multiply both sides of the above equation with the factor $2a\dot{a}$ and then integrate it using the continuity equation. This will lead to the modified Friedmann equation,
\begin{equation}\label{eq:FriedmannLovlock}
\sum_{i=1}^{m}\hat{c}_{i}\left(H^2 +\frac{k}{a^2}\right)^i =\frac{16\pi \ell_{\!\!_p}^{n-1}}{n(n-1)}\rho.
\end{equation} 

Gauss-Bonnet gravity is a special case of Lovelock gravity, where the summation in the Lovelock entropy (\ref{eq:entropy_lovlock}) takes only the first two terms, $m=2$. The entropy in Gauss-Bonnet gravity has the form  \cite{Sheykhi2013,PhysRevD.65.084014, PhysRevD.69.104025},
\begin{equation}\label{eq:entropy_gaussbonnet}
S= \frac{A}{4\ell_{p}^{n-1}}\left(1 + \frac{n-1}{n-3}\frac{2\tilde{\alpha}}{\tilde{r}_{_A}^{2}}\right),
\end{equation}
where $\tilde{\alpha}=(n-1)(n-3)/2$. Correspondingly the expansion law in Gauss-Bonnet gravity takes the form,   
\begin{equation}\label{eq: expansionlaw_gaussbonnet}
\frac{dV}{dt} = \frac{\ell_{p}^{n-1}\tilde{r}_{_A}H}{\left[1+2\tilde{\alpha}\tilde{r}_{_A}^{-2}\right]}\left[N_{sur} -\epsilon N_{bulk}\right],
\end{equation}
where surface degrees of freedom will be
\begin{equation}\label{eq:nsur_gaussbonnet}  
N_{sur} = \dfrac{A}{\ell_{p}^{n-1}}\left[1+ \tilde{\alpha}\tilde{r}_{\!\!_A}^{-2}\right].
\end{equation} 
The expansion law in Gauss-Bonnet gravity (\ref{eq: expansionlaw_gaussbonnet}) will lead to the modified Friedmann equation as a special case of the Friedmann equation in Lovelock gravity (\ref{eq:FriedmannLovlock}) as
\begin{equation}\label{eq:FriedmannGauss-Bonnet}
H^2 +\frac{k}{a^2}+\tilde{\alpha}\left(H^2 +\frac{k}{a^2}\right)^{2} =\frac{16\pi \ell_{\!\!_p}^{n-1}}{n(n-1)}\rho.
\end{equation} 
Hence the general expansion law can be easily reduced to more general gravity theories like Gauss-Bonnet and Lovelock gravity, and we have shown that the general expansion law can reproduce the modified Friedmann equations in the respective theories of gravity which shows the consistency and the general nature of the expansion law, we have derived.

In literature, R. G. Cai also extended the expansion law to Gauss-Bonnet and Lovelock gravity in flat FRW universe \cite{cai} and is further extended to non-flat FRW universe by A. Sheykhi \cite{Sheykhi2013}, 
 where they introduced the effective area $A_{eff}= 4\ell_{\!\!_p}^{n-1}S$ for defining the emergence of space, $\frac{dV_{\!\!_{eff}}}{dt}= \frac{\tilde{r}_{A}}{n-1} \frac{dA_{\!\!_{eff}}}{dt}$. In contrast to this, in using the general law of expansion, one need have to take any such arbitrary definitions either for area or rate of change of volume. Moreover, the degrees of freedom both on surface and bulk still contain the extra coefficient $\alpha$  for $n\ge 4$ as in the earlier extensions.
 By taking account of the arbitrariness in the definition of area in the work of Cai and Sheykhi, an alternative extension is tried in \cite{FARAGALI}, but it shows inconsistency with the Friedmann equations in Gauss-Bonnet gravity, which leads to more complications. 
\subsection{Expansion law in Horava-Lifshitz gravity}

In the case of Horava-Lifshitz gravity \cite{Horova1,Horova2,Horova3}, the entropy is in the form  \cite{Horova_entropy},
\begin{equation}\label{eq:entropy_Horava-Lifshitz}
S= \frac{A}{4\ell_{p}^{n-1}}+ \frac{\pi}{\omega}\ln\left(\frac{A}{4\ell_{p}^{(n-1)}}\right),
\end{equation}
having an additional logarithmic correction term, which vanishes for $\omega \to \infty $, and the entropy reduces to that in Einstein gravity. The time derivative of the above entropy is $\frac{(n-1)}{4}\left[\frac{1}{\ell_{p}^{n-1}\tilde{r}_{_A}}+\frac{4\pi}{n\omega V}\right]\frac{dV}{dt}.$
Then following the general expansion law in equation (\ref{eq:expansionlaw}), the expansion law in  Horava-Lifshitz gravity can be obtained as,  
\begin{equation}\label{eq: expansionlaw_Horava-Lifshitzt}
\frac{dV}{dt} = \frac{n\ell_{p}^{n-1} \tilde{r}_{_A}H V\omega}{\left[n\omega V + 4\pi \ell_{p}^{n-1}\tilde{r}_{_A}\right]}\left[N_{sur}-\epsilon N_{bulk}\right].
\end{equation}
Here, the surface degrees of freedom, $N_{sur}$ can be obtained using Eq. (\ref{eq:Nsur}) as,
\begin{equation}\label{eq:nsur_Horava-Lifshitz}  
N_{sur} = \dfrac{A}{\ell_{p}^{n-1}}+ \frac{8\pi}{(n+1)\omega },
\end{equation}
and the bulk degrees of freedom, $N_{bulk}$ is as given in Eq. (\ref{eq:Nbulk}). Next, we derive the modified Friedmann equation in Horava-Lifshitz gravity from the expansion law. Substituting for $V, A, E_{_{Komar}}$ and $T$ in the expansion law (\ref{eq: expansionlaw_Horava-Lifshitzt}), multiplying both sides by $2a\dot{a}$, along with little manipulations and with the use of continuity equation, the above expansion law can be expressed as 
\begin{equation}
\frac{d}{dt}\left(a^2 \tilde{r}_{_A}^{-2}\right) +\frac{8\pi\ell_{\!\!_p}^{n-1}}{n(n+1)\omega \Omega_{n}} \frac{d}{dt}\left(a^2\tilde{r}_{_A}^{-\left(n+1\right)}\right) =\frac{16\pi \ell_{\!\!_p}^{n-1}}{n(n-1)}\frac{d}{dt}\left(\rho a^{2}\right).
\end{equation}
Integrating the above equation, we get
\begin{equation}
\tilde{r}_{_A}^{-2} +\frac{8\pi\ell_{\!\!_p}^{n-1}}{n(n+1)\omega \Omega_{n}} \tilde{r}_{_A}^{-\left(n+1\right)} =\frac{16\pi \ell_{\!\!_p}^{n-1}}{n(n-1)}\rho
\end{equation}
or
\begin{equation} \label{eq:Friedmanneqn_horava}
H^{2}+\frac{k}{a^2} +\frac{8\pi\ell_{\!\!_p}^{n-1}}{n(n+1)\omega \Omega_{n}} \left(H^{2}+\frac{k}{a^2}\right)^{\frac{\left(n+1\right)}{2}} =\frac{16\pi \ell_{\!\!_p}^{n-1}}{n(n-1)}\rho.
\end{equation}
In general, the above equation (\ref{eq:Friedmanneqn_horava}), which is obtained from our expansion law is the modified Friedmann equation in the Horava-Lifshitz gravity. For $ (3+1) $ FRW universe, this will reduce to
\begin{equation} \label{eq:Friedmanneqn_horava_3+1}
H^{2}+\frac{k}{a^2} +\frac{1}{2\omega} \left(H^{2}+\frac{k}{a^2}\right)^{2} =\frac{8\pi \ell_{\!\!_p}^{2}}{3}\rho.
\end{equation}
For $ \omega \to \infty $, the modified Friedmann equation will reduce to the standard Friedmann equation in Einstein gravity.
 
\subsection{Expansion law for non-extensive entropy} 
The general expansion law exclusively depends on the form of entropy, which is different in different gravity theories. The general expansion law can thus be used to formulate the expansion law for any modified entropy. For illustration, we formulate the expansion law for Tsallis entropy \cite{Tsallis}, given as,
\begin{equation}\label{eq:Tsallis_entropy}
S = \frac{A_{0}}{4 \ell_{\!\!_p}^{n-1}} \left(\frac{ A}{A_{0}}\right)^{\delta} ,
\end{equation}
where $A_{0}$ is a constant with the dimension of the area, and $ \delta $ is a dimensionless constant.   
The time derivative of the above entropy is $\frac{(n-1)\delta A^{\delta -1}}{4 \ell_{\!\!_p}^{n-1} \tilde{r}_{\!\!_A}A_{0}^{\delta-1}}\frac{dV}{dt}.$ The corresponding surface degrees of freedom can then be obtained using Eq. (\ref{eq:Nsur}) as,
\begin{align}\label{eq:nsur_nonext}  
N_{sur} &= -\frac{2}{(n-1)}4 \tilde{r}_{_A}^{n+1}\int \frac{1}{\tilde{r}_{_A}^{n+1}}dS 
\nonumber\\
&=\frac{2\delta}{\left((n+1)-\delta(n-1)\right)} \frac{A^{\delta}}{\ell_{\!\!_p}^{n-1}A_{0}^{\delta - 1}},
\end{align}
 and $N_{bulk}$ can be taken as given in Eq. (\ref{eq:Nbulk}).  The generalized expansion law 
with Tsallis entropy  can be obtained as,  
\begin{equation}\label{eq:Tsallis_expansionlaw}
\frac{dV}{dt} =\frac{A_{0}^{\delta-1} }{\delta A^{\delta -1}}\ell_{\!\!_p}^{n-1}\tilde{r}_{_A}H\left[N_{sur} -\epsilon N_{bulk}\right].
\end{equation}
Substituting for $V, A, E_{_{Komar}}$ and $T$ in the expansion law (\ref{eq:Tsallis_expansionlaw}), multiplying both sides by $ 2a\dot{a}  $, and by using the continuity equation, one gets
\begin{align} \label{lawTsallis}
\frac{d}{dt}&\left(a^{2}\tilde{r}_{\!\!_A}^{(n-1)\delta- (n+1)}\right) = \\
&\quad \frac{\left[(n+1) - (n-1)\delta\right]  A_{0}^{\delta-1}}{\delta \left(n\Omega_{n}\right)^{\delta-1}} \frac{8\pi\ell_{\!\!_p}^{n-1}}{n(n-1) }\frac{d}{dt}\left(\rho a^2\right).
\end{align}
The integral of the above equation is
\begin{equation}\label{frwTsallis}
\left(H^2 +\frac{k}{a^2}\right)^{\frac{(n+1) - (n-1)\delta }{2}} = \frac{\left[(n+1) - (n-1)\delta\right]  A_{0}^{\delta-1}}{\delta \left(n\Omega_{n}\right)^{\delta-1}} \frac{8\pi\ell_{\!\!_p}^{n-1}}{n(n-1) }\rho .
\end{equation}
This is the modified Friedmann equation that emerged from the general expansion law corresponding to a Tsallis entropic modified gravity. This reduces to the standard Friedmann equation in Einstein gravity when both $\delta$ and $A_{0}$ are equal to one. For $n=3$, the law of emergence in equation (\ref{lawTsallis}) and the corresponding Friedmann equation in (\ref{frwTsallis}), will reduce to that obtained by Sheykhi \cite{SHEYKHI2018118}, for Tsallis entropy in (3+1) dimensions. It is possible to extend this procedure to other general entropies like Barrow entropy, Renyi entropy etc.

\section{Significance of the unified expansion law }\label{sec.significance}

 In the present paper, we realize a unified expansion law, Eq. (\ref{eq:expansionlaw}), from the first law of thermodynamics. In reference \cite{Krishna2022}, we establish the connection of the law of emergence (and its prior generalizations) with other well-established results in thermodynamics. It was shown that the expansion law in references \cite{cai,Sheykhi2013,yang2012} could be derived from the unified first law of thermodynamics using modified Friedmann equations in each gravity theory. It was also shown that the expansion law predicts the evolution of the universe towards a state of maximum horizon entropy. Further, it was concluded that the first law of thermodynamics, along with the additional constraints imposed by the maximization of the horizon entropy, can together lead to the law of emergence. In the present manuscript, motivated by these results, we realize a unified expansion law entirely in terms of thermodynamic variables, which can be used in general gravity theories. The aim of the present paper is different from the results in \cite{Krishna2022}. The present one suggests a form invariant unified expansion law for a general set of gravity theories, whereas the last one studies the connection between thermodynamic principles and the expansion law.
 
In reference \cite{Mahith2018}, the authors derived the various forms of the already proposed expansion law from the first law of thermodynamics using Friedmann equations in the respective gravity theories. In the present paper, we derive a general expansion law from the first law of thermodynamics in a more general approach, and we never used the Friedmann equations in between steps. Hence the present derivation gives a more general form for the expansion law in terms of entropy instead of effective area/volume, which can be used to obtain the expansion law for any modified gravity theory. The derivation also provides an alternative definition for surface degrees of freedom that can be used in a general set of gravity theories. 
 
The surface degrees of freedom in the emergent gravity paradigm is generally assumed as $N_{sur}=4S$ \cite{PhysRevD.81.124040,Padmanabhan:2003pk,Chakraborty:2014rga}. However, one can note that $N_{sur}$ used in the previous generalizations of the expansion law are not equal to $4S$  for more general theories like Gauss-Bonnet and Lovelock gravity \cite{cai,Sheykhi2013}. In our present paper, we showed that the surface degrees of freedom in general gravity theories is not equal to $4S$. This happens naturally when we derive the expansion law from the first law of thermodynamics. 
 In Einstein's gravity, the equipartition energy reduces to $2TS$, and the surface degrees of freedom will take the form $N_{sur}=4S$. 
 
One can ask whether the unified formalism for the law of expansion, which we have formulated for general gravity theories, could be applied to cosmological models like the standard  $ \Lambda CDM $ model and  the dynamical dark energy models. It has to be noted that our unified law of expansion will reduce to the law of expansion proposed by Padmanabhan in (3+1) Einstein gravity. In 
one of 
our earlier works, we have explicitly proved the consistency of the standard $ \Lambda CDM $ model with the law of expansion \cite{krishna1}. In a subsequent work \cite{Krishna_MPLA},  
we have applied Padmanabhan's law of emergence to the dynamical dark energy models by suitably defining the degrees of freedom on the surface and in bulk and have 
shown that the dynamical dark energy models, which evolve to a final de Sitter phase  
are consistent with the law of expansion. Since the running vacuum models \cite{running_qft1,running_qft2,Sola:2015,Sola:2016,SolaPeracaula:2022hpd,SolaPeracaula:2019TD_interpretations}, which can be considered as the potential candidates to cure the current tensions in the $ \Lambda CDM $ \cite{SolaPeracaula:2021}, are  
dynamical dark energy models 
with a final de Sitter epoch,  
it is evident that they will  
be consistent with the law of expansion. Thus, it is important to note that our unified law of expansion  
is consistent with 
the prominent models of cosmology, like standard $\Lambda CDM$ and  
the running vacuum models.


\section{Discussions}\label{sec.4}

The law of expansion, which describes the universe's evolution as the emergence of space, was proposed by Padmanabhan for a flat universe in the context of Einstein's gravity. Many have extended this to higher dimensional gravity theories in different ways, both for the flat and non-flat universe. In this manuscript, we have derived a general, unified law of expansion using the first law of thermodynamics.  The advantage of the unified law (\ref{eq:expansionlaw}) is that both the emergence of space, that is, the time evolution of the horizon volume and the degrees of freedom on the horizon surface are being expressed in terms of the entropy. Since entropy is unique to each gravity theory, one can easily get the law of expansion in a given gravity theory by substituting the corresponding form of entropy. Hence one need not have to define any effective volume, or modified surface degrees of freedom, as employed in many previous attempts to define the law of emergence. Instead, here we naturally get the law of expansion by using the entropy corresponding to the gravity theory. This general nature of the unified law of expansion comes from the first law of thermodynamics.
 
In general, we have derived the unified expansion law for a general (n+1) dimensional FRW universe. For a flat (3+1) dimensional FRW universe, the unified expansion law in Einstein's gravity and corresponding degrees of freedom $N_{sur} $ and $N_{bulk}$ precisely gives the expansion law proposed by Padmanabhan and the respective degrees of freedom \cite{paddy2012jun}. This shows the consistency of our unified expansion law with the original proposal. 
  	
In the original proposal, Padmanabhan considered the temperature of the horizon as $T=H/2\pi,$ the Gibbons-Hawking temperature, to obtain the bulk degrees of freedom, $N_{bulk},$ instead of taking as $\mathcal{T}=\kappa/2\pi $ (\ref{eq:temperature}). It is to be noted that, for the de Sitter horizon, $\mathcal{T}$ becomes equal to $T.$ To derive the law of expansion, we have started with the unified first law of thermodynamics with temperature $\mathcal{T}.$  In this method, we have to extract the bulk degrees of freedom by retaining temperature as $\mathcal{T}.$  While doing so, we found that the bulk degrees of freedom arose as $N_{bulk}= E_{Komar} (1-\epsilon_T)/((1/2)\mathcal{T}),$ which seems to deviate from the standard form of equipartition rule, by a factor $(1-\epsilon_T).$  However it turns out that $\mathcal{T}/(1-\epsilon_T)$ is exactly equivalent to $T,$ the Gibbons-Hawking temperature. Hence the temperature $T$ has been naturally selected here for defining the bulk degrees of freedom. 
In reference \cite{DavidIvanBooth}, the confusion regarding the choice of apparent horizon temperature is studied, they also favours the temperature $T= 1/2\pi \tilde{r}_{_A}$ rather than $\mathcal{T}=\kappa/2\pi$. 

Further, the emergence of the spacial volume in the unified expansion law is written in terms of $\frac{dS}{dt}$ rather than  $\frac{dV}{dt}$. Hence it  naturally selects what is known as areal volume instead of proper invariant volume, eliminating the discrepancy in the use of proper invariant volume. There is confusion in choosing the volume of the horizon while formulating the expansion law.
The expansion law in a non-flat universe can not be properly formulated using the proper invariant volume, but it can only be done using the areal volume. This is discussed in reference \cite{Hareesh_2019, Chang-Young:2013gwa}. In our derivation, since it is written directly in terms of entropy, Eq. (\ref{eq:expansionlaw}) naturally selects the areal volume rather than the proper invariant volume and avoids such issues.

From the general expansion law (\ref{eq:expansionlaw}), we have obtained the expansion law corresponding to different theories of gravity like (n+1) Einstein, Lovelock, Gauss-Bonnet and Horava-Lifshitz. In deriving the law of expansion in these gravity theories, we need not have to define any effective volume or surface degrees of freedom like in the previous extensions of the law\cite{cai,Sheykhi2013,SHEYKHI2018118}. Instead, just a substitution of the corresponding expression for entropy in the general law is sufficient. We have also used the general law to obtain the law of expansion with non-extensive entropy, like Tsallis entropy. 

We compared the general expansion law with the previously proposed laws for different gravity theories. The advantage of our approach is that both $N_{sur}$ and $N_{bulk}$ are naturally identified from the first law of thermodynamics. On the other hand, in the previous proposals, authors have adopted ad hoc definitions for these degrees of freedom. Hence the unified law of expansion proposed by us, following the thermodynamic principles, is general and simpler than the previous proposals. The general formalism we have proposed in equation (\ref{eq:expansionlaw}) can be applied to any theory of gravity. 

The important point is that the derivation of the expansion law from the first law of thermodynamics also gives an alternative definition for the surface degrees of freedom, $N_{sur}$, which is given in Eq. ($ \ref{eq:Nsur} $). 
 It is also shown that this alternative definition is valid for a more general set of gravity theories, including Lovelock gravity, for which the entropy is not proportional to the horizon area. Our alternative definition of $N_{sur}$ is similar to the definitions of $N_{sur}$ in the previous expansion laws in Gauss-Bonnet, and Lovelock gravity \cite{cai,Sheykhi2013}, except the appearance of a coefficient $\alpha=\frac{(n-1)}{2(n-2)} $, where $ n $ is the spatial dimension. The additional coefficient $\alpha $ in the literature is motivated by Verlinde's paper \cite{Verlinde}, where the author assumes an extra factor $1/\alpha $ with Newton's constant in higher dimensional space. Also, the minimal area $\ell_{p}^{2}$ for (3+1) dimensional universe is generalized as $\ell_{p}^{n-1}/\alpha$ for the (n+1) dimensional universe \cite{Chang-Young:2013gwa}. Correspondingly, the definition for bulk degrees of freedom also gets an additional factor, $\alpha$ in these formulations \cite{caiEntropicForce}. However, in the emergent gravity paradigm, the minimal area in (n+1) dimensional space-time is defined as $\ell_{p}^{n-1}$ in a quiet natural way \cite{Chakraborty:2014rga,Chakraborty:2014joa}. We follow the simple generalization without the additional factor $\alpha$. 
This approach, the formulation of the expansion law without $\alpha$ is more appealing and useful in further studies on the expansion law in more general contexts of emergent gravity paradigm.            
	
In the emergent gravity paradigm,  the equipartition energy of the horizon surface generally assumes the form $E_{eq}=2TS$. Correspondingly, the surface degrees of freedom will take the form,  $N_{sur}=2TS/(1/2k_{B}T) =4S$ \cite{PhysRevD.81.124040,Padmanabhan:2003pk,Chakraborty:2014rga}. However, our derivation of the expansion law from the first law of thermodynamics shows that the equipartition energy on the horizon surface can be identified as $ E_{eq}=2V \int \frac{-TdS}{V}$. Corresponding to this, we get the surface degrees of freedom different from $4S$ in more general theories of gravity. What will be the key reason for this? Is there any alternative formalism for expansion law that has taken care of such thermodynamic relations? These are the questions which we have to investigate further.

\acknowledgments
 
V. T. H. B. acknowledges	Cochin University of Science and Technology for the Research Fellowship. P. B. K. acknowledges Cochin University of Science and Technology for the Postdoctoral Fellowship.

%

\end{document}